\documentclass[preprint,onecolumn,nofootinbib]{revtex4}
\pdfoutput=1
\usepackage{float}
\usepackage[colorlinks=true,linkcolor=blue,urlcolor=blue,filecolor=black,citecolor=red,pdfstartview=FitV,pdftitle={},pdfsubject={},pdfkeywords={},pdfpagemode=None,bookmarksopen=true]{hyperref}
\usepackage{graphicx}%Include figure files
\usepackage{amsmath}
\usepackage{amsfonts}
\usepackage{amssymb,ulem}
\usepackage{color,xcolor}%
\usepackage{CJK}
\usepackage{subfigure}

\usepackage{dcolumn}% Align table columns on decimal pointl
\newcommand{\f}{\begin{equation}}
\newcommand{\ff}{\end{equation}}
\newcommand{\fa}{\begin{eqnarray}}
\newcommand{\ffa}{\end{eqnarray}}

\begin{document}

\title{Transverse Goldstone mode in holographic fluids with broken translations}

\author{Yuan-Yuan Zhong$^{1}$}
\thanks{zhongyuanyuan@mail.dlut.edu.cn}
\author{Wei-Jia Li$^{1}$}
\thanks{weijiali@dlut.edu.cn} \affiliation{$^1$ Institute of Theoretical Physics, School of Physics, Dalian University of Technology, Dalian 116024, China.
}

\begin{abstract}
In this paper we investigate the low energy shear modes in fluid systems with spontaneously broken translations by a specific holographic model. In absence of momentum relaxation, we find that there exist two decoupled gapless modes in the transverse channel, one of which is purely diffusive and the other corresponds to vortex like excitations. The diffusive mode is associated with the conservation of momentum and the vortex mode can be viewed as the Goldstone mode of the spontaneous symmetry breaking. Switching on an external source which breaks the translations explicitly but weakly, the would-be gapless modes both get relaxed and acquire a tiny mass gap. Finally, in the strong momentum relaxation regime, we find a (pseudo-)diffusive-to-sound crossover that is set by a momentum gap.
\end{abstract}
\maketitle
\tableofcontents
\section{Introduction}
In modern physics, symmetry principles play a crucial role in understanding laws of nature. Using the language of effective field theory (EFT), theoretical physicists have been attempting to classify and describe various different phases of condensed matter systems in terms of symmetry and symmetry breaking patterns during past few decades \cite{Nicolis:2013lma,Nicolis:2015sra}. Such an EFT treatment has not only resulted to be very convenient for applications but also provided new insights into some widely-observed phenomena and deepened our understanding about them, for instance, the relationship between the rigidity of solids and the spontaneous breaking of spatial translations, the Goldstone nature of acoustic phonons, etc \cite{Leutwyler:1996er,Alberte:2018doe}. Nevertheless, broken translations should not be viewed as a sufficient criterion for distinguishing solids from fluids in which there 
do not exist transverse phonons (propagating shear sound modes) due to the lack of rigidity.\footnote{This statement works for low energy and long distance. However, shear propagating modes may still survive at certain short distance in fluids and the related momentum scale is called the momentum gap or ``$k$-gap". For a detailed discussion on this, one refers to \cite{Baggioli:2019jcm}.} One can manifestly show  by EFT that if the system has a larger symmetry group (called the volume preserving diffeomorphisms), it can no longer resist shear deformations or mediate transverse sounds \cite{Nicolis:2013lma,Nicolis:2015sra}. Then, questions arise that what is the difference between a fluid with spontaneously broken translations and a regular one that is translationally invariant? Since transverse sounds do not exist in the former either, what is destiny of the Goldstone modes in the transverse channel? 

In this paper, we will address these questions by using the holographic method \cite{Maldacena:1997re,Aharony:1999ti,Ammon:2015wua,Zaanen:2015oix,Hartnoll:2016apf,Baggioli:2019rrs}. One of the biggest advantages of holography is that the dissipative effects at finite temperature can be simply resolved by black holes in one dimension higher \cite{Policastro:2001yc}. Borrowing ideas from the EFT formulation, one can now realize  breaking of translations in isotropic and homogeneous holographic models by imposing certain global internal symmetries and space-dependent configurations about several scalars in the bulk which violates the diffeomorphism invariance and gives gravitons an effective mass \cite{Vegh:2013sk,Blake:2013owa}. During the past few years, depending upon different setups, it has been widely investigated how the explicit breaking of translations affects the transport properties in homogeneous holographic systems \cite{Andrade:2013gsa,Donos:2013eha,Donos:2014cya,Gouteraux:2014hca,Davison:2014lua,Kim:2014bza,Kim:2015dna,Jeong:2017rxg,Baggioli:2014roa,Baggioli:2015zoa,Alberte:2015isw,Baggioli:2016oqk,Amoretti:2016cad,Ling:2015epa,Ling:2016ien,An:2020tkn,Amoretti:2021lll}. And more recently,  it was found that even the spontaneously broken translations can be
realized by choosing different UV boundary conditions in the same model which makes it possible to introduce the phonon dynamics on boundary and compare the results from holography with the experiments of cuprates \cite{Donos:2018kkm,Donos:2019tmo,Donos:2019txg,Donos:2019hpp,Amoretti:2016bxs,Amoretti:2017frz,Amoretti:2017axe,Amoretti:2018tzw,Amoretti:2019cef,Amoretti:2019kuf,Alberte:2017oqx,Baggioli:2018vfc,Ammon:2019wci,Baggioli:2019aqf,Baggioli:2019mck,Amoretti:2019buu,Amoretti:2020mkp,Baggioli:2020edn,Baggioli:2020qdg,Pan:2021cux,Baggioli:2021tzr,Amoretti:2021fch,Wang:2021jfu,Arean:2020eus,Wu:2021mkk}. The simplest model of this type is the so-called holographic axion model in which the scalar sector satisfies an internal Euclidean symmetry (see \cite{Baggioli:2021xuv} for the most up-to-data review on this specific model). 
The discussions of this work will be based on a holographic charged fluid model where the spontaneously broken translations are realized by an axion-gauge coupling. It was introduced as an effective model to describe momentum relaxation effects at the beginning \cite{Gouteraux:2016wxj,Baggioli:2016pia}. However, the correct interpretation pointing to the spontaneous symmetry breaking(SSB) was unveiled in a later work \cite{Li:2018vrz}. Since the longitudinal low energy excitations in fluids are not quite different from those in solids, we will focus only on the transverse channel in this work. 

The contents of the paper are organized as follows: In section \ref{section2}, we introduce the holographic bulk action, obtain the black hole solution and list two possible ways of breaking translations in this model, explicit and spontaneous. In section \ref{section3}, the transverse low energy modes are investigated both in the purely SSB case and in the pseudo-spontaneous breaking case by computing the quasi-normal modes (QNMs) of the black hole numerically. In section \ref{section4}, we conclude.

\section{Holographic setup}\label{section2}
In holography, fluids at finite density can be described by charged black hole geometries with one extra dimension. The simplest case is the Reisner-Nordstr{\"o}m(RN) black hole which can be achieved from the Einstein-Maxwell action,
\begin{eqnarray}\label{action1}
\mathcal{S}_{\text{EM}}=\int d^{4} x \sqrt{-g}\left(R-2 \Lambda-\frac{1}{4} F^{2}\right),
\end{eqnarray}
where $R$ is the Ricci scalar, $\Lambda$ is the negative cosmological constant which for simplicity is set as $\Lambda=-3$ so that the AdS radius is normalized and $F^{2}=F_{\mu \nu} F^{\mu \nu}$ with $ F_{\mu \nu} \equiv \nabla_{\mu} A_{\nu}-\nabla_{\nu} A_{\mu}$, strength of the Maxwell field that introduces a finite charge density on boundary. 

In order to break the translations of the boundary system while retaining homogeneity, we couple two massless scalars $ \phi^I, \, I=x,y$ (which are called axions) to the Einstein-Maxwell sector. This idea is quite similar as introducing the co-moving coordinates as dynamical fields and imposing an internal shift symmetry in the EFT formulation of solids as well as fluids \cite{Nicolis:2013lma,Nicolis:2015sra,Alberte:2018doe}. Then, at the lowest order in derivatives, the terms containing axions should be built in terms of 
\begin{eqnarray}
\frac{1}{2}\partial^{\mu} \phi^{I} \partial_{\nu} \phi^{J}.
\end{eqnarray}
We expect that the dual system is a fluid and define the kinetic term of axions as follows, 
\begin{eqnarray}
Z =\frac{1}{2} \left(\partial^\mu\phi^I\partial_\mu\phi^I\partial^\nu\phi^J\partial_\nu\phi^J-\partial^\mu\phi^I\partial_\mu\phi^J\partial^\nu\phi^I\partial_\nu\phi^J\right).
\end{eqnarray}
Note that this term is just the determinant of the matrix $\frac{1}{2}\partial^{\mu} \phi^{I} \partial_{\mu} \phi^{J}$, which satisfies the volume perserving diffeomorphisms \cite{Alberte:2015isw}. Without loss of generality, one can also consider the axion-gauge interactions that takes the form\footnote{Unlike another kind of coupling $\text{Tr}[\mathcal{X}]F^2$, this term does not contribute a finite shear modulus to the boundary system in the absence of external magnetic field. Therefore, the boundary system is always a fluid.}
\begin{eqnarray}
\operatorname{Tr}\left[\mathcal{X} F^{2}\right] \equiv {\mathcal{X}^{\mu}}_{\nu} {F^{\nu}}_{\rho} {F^{\rho}}_{\mu}\qquad \text{with} \qquad {\mathcal{X}^{\mu}}_{\nu}=\frac{1}{2} \sum_{I= x,y} \partial^{\mu} \phi^{I} \partial_{\nu} \phi^{I}.
\end{eqnarray}
Then, the bulk terms containing axions can be constructed as 
\begin{eqnarray}
\mathcal{S}_{\text{A}}=\int d^{4} x \sqrt{-g}\left(\lambda Z+\frac{\mathcal{J}}{4}\operatorname{Tr}\left[\mathcal{X} F^{2}\right]\right),
\end{eqnarray}
 where we require that the coupling constants $\lambda \geq 0$  and $0 \leq \mathcal{J} \leq 2/3$ so that the dual system avoids the problem of ghost \cite{Baggioli:2014roa}, and obeys unitarity as well as causality \cite{Gouteraux:2016wxj}. 

In this model, there exists a homogeneous and isotropic black brane solution which in Eddington-Finkelstein (EF) coordinates can be expressed as
\begin{align}
&d s^{2}=\frac{1}{u^{2}}\left[-f(u) d t^{2}-2 d t d u+d {x^i}^{2}\right],\nonumber\\
&f(u)=1-\frac{u^{3}}{u_{h}^{3}}+\frac{1}{8} u^{4} \alpha^{4} \lambda-\frac{1}{8} u^{3} u_{h} \alpha^{4} \lambda+\frac{u^{4} \mu^{2}}{4 u_{h}^{2}}-\frac{u^{3} \mu^{2}}{4 u_{h}},
\end{align}
and 
\begin{eqnarray}\label{solution2}
&&A=A_{t}(u)\,d t, \qquad \text{with} \qquad A_{t}=\mu-\rho\,u\\
\label{solution3}&&\phi^{I}= \alpha \delta_{i}^{I} x^{i},
\end{eqnarray}
where $u=0$ and $u=u_h$ denote the boundary and horizon respectively,
$\mu$ and $\rho =\mu / u_{h}$ correspond to the chemical potential and charge density on boundary.  Note that the bulk axions are dual to boundary scalar operators $\mathcal{O}^I$. Their space dependence hence implies the breaking of translations. And the constant $\alpha$ in (\ref{solution3}) is a parameter characterizing the strength of the symmetry breaking. The black hole temperature and entropy density are given by
\begin{equation}
T=\frac{24-u_{h}^{4} \alpha^{4} \lambda-2 u_{h}^{2} \mu^{2}}{32 \pi u_{h}},\qquad s=\frac{4\pi}{u_h^2},
\end{equation}
which should be identified as the temperature and entropy density of the dual system.

As was pointed in previous work \cite{Li:2018vrz}, near the AdS boundary $u=0$, the general solution of the scalar equations can be expanded as 
\begin{equation}
\phi^{I}(u,t,x^i)=\phi_{I}^{(0)}(t,x^i)+\phi_{I}^{(1)}(t,x^i)u+\dots   \qquad  \text{for}  \qquad   \lambda\neq0, \mathcal{J}=0
\end{equation}
and 
\begin{equation}
\phi^{I}(u,t,x^i)=\phi_{I}^{(-1)}(t,x^i) u^{-1}+\phi_{I}^{(0)}(t,x^i)+\dots \qquad  \text{for} \qquad   \lambda = 0, \mathcal{J}\neq0.
\end{equation}
Then, in the first case above, the special solution (\ref{solution3}) is the leading behavior of the UV expansion and plays the role of external source of the operator $\mathcal{O}^I$, following the standard quantization. On the contrary, the same solution becomes subleading  in the second case and now corresponds to a non-trivial condensate
$\langle\mathcal{O}^I\rangle \sim \alpha \delta _i^I x^i$  which means the breaking of translations is spontaneous. In consequence, this model can easily realize different ways of the symmetry breaking via different settings on the bulk couplings.

Next, we turn to the low energy spectrum that can be achieved by investigating perturbations upon the background, $g_{\mu \nu}=\bar{g}_{\mu \nu}+\delta g_{\mu \nu}, A_{\mu}=\bar{A}_{\mu}+\delta A_{\mu}$ and $\phi^{i}=\bar{\phi}^{i}+\delta \phi^{i}$, where the quantities with bars are the background values. Without loss of generality, we choose the momentum $k$ to be parallel to the $y$-axis. Then, the transverse modes are associated with
\begin{align}
&\delta g_{t x}=\frac{1}{u^{2}} h_{t x}(u) e^{-i \omega t + i k y}, 
 \quad \delta g_{x y}=\frac{1}{u^{2}} h_{x y}(u) e^{-i \omega t  + i k y},\nonumber\\&
\quad \delta A_{x}=a_{x}(u) e^{-i \omega t  + i k y}, \quad \delta \phi^{x}=\phi_{x}(u) e^{-i \omega t  + i k y} .
\end{align}
For convenience, assuming the radial gauge, i.e. $\delta g_{x u}=0$, we obtain following linearized equations, 
\begin{align}\label{leom1}
0=&-2 \mathcal{J} u \alpha \mu^{2} h_{t x} + 2 i \mathcal{J} u \alpha \mu \omega a_{x} -2 i \mathcal{J} u \mu^{2} \omega \phi_{x} - 2 \alpha^{3} \lambda h_{ t x}^{\prime} - \mathcal{J} u^{2} \alpha \mu^{2} h_{ t x}^{\prime}-\nonumber\\ 
& 4i\alpha^{2} \lambda \omega \phi_{x}^{\prime}-2 i \mathcal{J} u^{2} \mu^{2} \omega \phi_{x}^{\prime} -2 \mathcal{J} u \mu^{2} f \phi_{x}^{\prime} u - \alpha^{2} \lambda f^{\prime} u \phi_{x}^{\prime} u -\mathcal{J} u^{2} \mu^{2} f^{\prime} u \phi _{x}^{\prime} u -\nonumber\\&\left(2 \alpha^{2} \lambda+\mathcal{J} u^{2} \mu^{2}\right) f  \phi_{x}^{\prime \prime},\\ \label{leom2}
0=&-2 i k h_{t x} - 2 i \omega h_{x y} + i k u h_{t x}^{\prime} + \left [-2 f + u\left(2 i \omega+f^{\prime}\right)\right] h_{x y}^{\prime}+u f h_{x y}^{\prime \prime},\\  \label{leom3}
0=&2 \mathcal{J} u \alpha^{2} \mu h_{t x} -4 k^{2} a_{x} + 2 \mathcal{J} u \alpha^{2}\left(k^{2} u - i \omega\right) a_{x} + 2 i \mathcal{J} u \alpha \mu \omega \phi_{x} - 4 h_{t x}^{\prime}\nonumber\\
& + \mathcal{J} u^{2} \alpha^{2} h_{t x}^{\prime} + 8 i \omega a_{x}^{\prime}- 2 i \mathcal{J} u ^{2} \alpha^{2} \omega a_{x}^{\prime} - 2 \mathcal{J} u \alpha^{2} f a_{x}^{\prime} + 4 f^{\prime} a_{x}^{\prime} \nonumber\\& - \mathcal{J} {u}^{2} \alpha^{2} f^{\prime} a_{x}^{\prime} + \left(4 - \mathcal{J} {u}^{2} \alpha^{2}\right) f a_{x}^{\prime \prime},\\  \label{leom4}
0=&i \left\{u^{2} \omega\left[\left(4-J u^{2} \alpha^{2}\right) \mu a_{x} + \alpha\left(2 \alpha^{2} \lambda + \mathcal{J} u^{2} \mu^{2}\right) \phi_{x}\right]-4 \omega h_{t x}^{\prime}-4 k f h_{x y}^{\prime}\right\} + \nonumber\\
&\left(4 k^{2}+2 u^{2} \alpha^{4} \lambda + \mathcal{J} u^{4} \alpha^{2} \mu^{2}\right) h_{t x} + 4 k \omega h_{x y} + u^{2} \alpha\left(2 \alpha^{2} \lambda+J u^{2} \mu^{2}\right) f \phi_x^{\prime}
,\\ \label{leom5}
0=&8 h_{t x}^ {\prime} +4 i k u h_{t x}^{\prime} - {u}^{3}\left[\left(-4 + \mathcal{J} u^{2} \alpha^{2}\right) \mu \delta_{x}^{\prime} + \alpha\left(2 \alpha^{2} \lambda+ \mathcal{J} {u}^{2} \mu^{2}\right) \phi_{x}^{\prime} \right] - 4 u h_{t x}^{\prime \prime}.
\end{align}
The UV expansions of $h_{t x}$, $h_{x y}$ and $a_{x}$ close to $u=0$ are given by
\begin{equation}
\begin{array}{l}
h_{t x}=h^{(0)}_{t x}(1+\ldots)+h^{(3)}_{t x} u^{3}(1+\ldots), \\
h_{x y}=h^{(0)}_{x y}(1+\ldots)+h^{(3)}_{x y} u^{3}(1+\ldots), \\
a_{x}=a^{(0)}_{x}(1+\ldots)+a^{(1)}_{x} u(1+\ldots),
\end{array}
\end{equation}
Retarded Green functions can be read as
\begin{align}
&\mathcal{G}^R_{T_{tx}T_{tx}}=\left(2\Delta-d\right)\frac{h^{(3)}_{tx}}{h^{(0)}_{tx}}=3\frac{h^{(3)}_{tx}}{h^{(0)}_{tx}},\\
&\mathcal{G}^R_{T_{xy}T_{xy}}=\left(2\Delta-d\right)\frac{h^{(3)}_{xy}}{h^{(0)}_{xy}}=3\frac{h^{(3)}_{xy}}{h^{(0)}_{xy}},\\
& \mathcal{G}^R_{J^xJ^x}=\frac{a^{(1)}_{x}}{a^{(0)}_{x}},
\end{align}
while the expression of $\mathcal{G}^R_{\mathcal{O}^{x}\mathcal{O}^{x}}$ depends on the boundary condition of axions in the UV. Then, all the correlators can be obtained once one solves (\ref{leom1})-(\ref{leom5}). It is important to point out that axions do not appear in (\ref{leom2}) explicitly. As a result of that the graviton $h_{xy}$ still remains massless and\footnote{The axion-gauge coupling modifies the equations of other components of gravitons. Then, it affects the thermoelectric transport in the dual system. For its impacts on the charge transport, one refers to \cite{Li:2018vrz}.}
\begin{align}
\mathcal{G}^R_{T_{xy}T_{xy}}\left(\omega,k=0\right)=-i\omega\frac{s}{4\pi}+\dots,
\end{align}
which implies that the shear modulus $G$ (which is defined as the real part of the stress tensor Green function at zero frequency) is vanishing. Therefore, the boundary system is indeed a fluid with the shear viscosity touching the celebrated KSS bound \cite{Kovtun:2004de,Alberte:2016xja}.

In general, the excitations can be extracted directly from poles of Green functions for finite frequency and momentum. In holography, these poles coincide with the QNMs of black holes \cite{Son:2002sd, Berti:2009kk}. This allows us to turn the bulk equations into secular equations with complex frequencies and real momenta, hence simplifying the numeric computation significantly.

\section{Low energy shear modes}\label{section3}
From now on, we will focus on the low energy modes in the transverse channel by computing the lowest-lying QNMs. We begin our discussions
with the purely SSB case, then turn to the case in the presence of an external source.
\subsection{Hydrodynamic mode and frozen Goldstone mode}\label{setionA}
\begin{figure*}[htbp]
\vspace{0.3cm}
\centering
\includegraphics[width=0.48\textwidth]{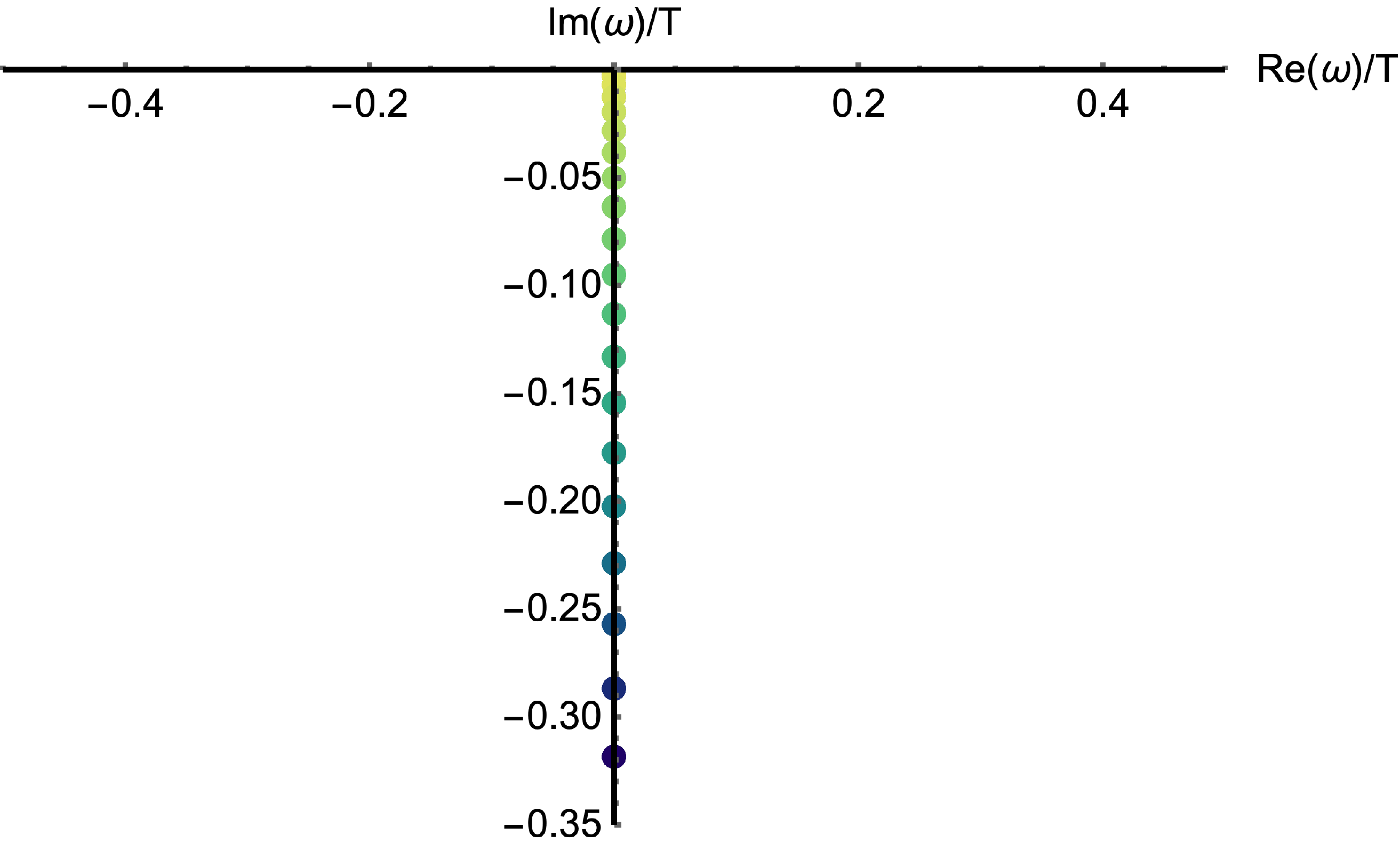}
\includegraphics[width=0.48\textwidth]{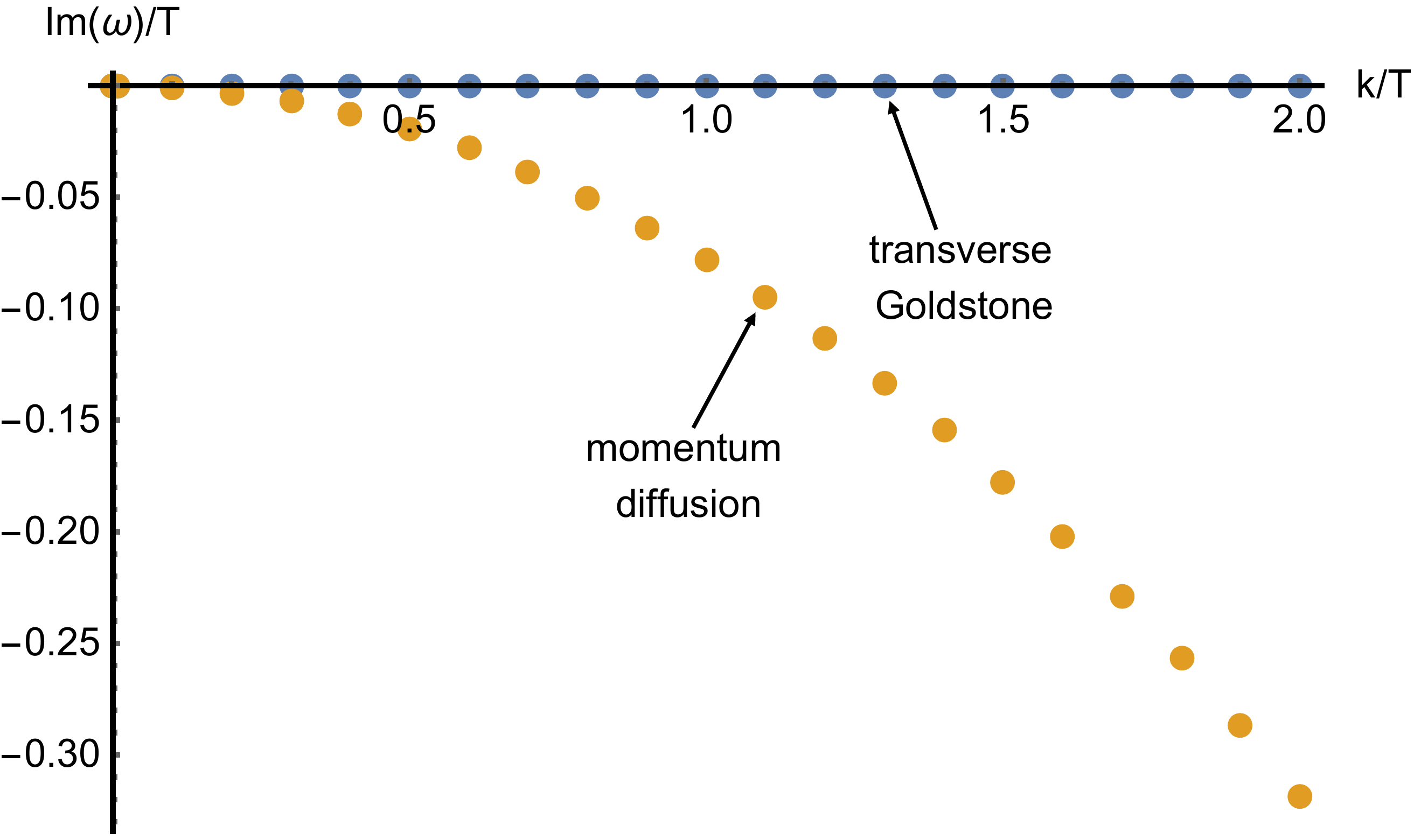}
\caption{The motion of the lowest-lying QNMs along the imaginary axis with the increase of $k$ by fixing $\lambda=0$ and $\mathcal{J}\neq0$. \textbf{Left}:  Imaginary part of the frequency of the QNMs with fixed $T/\mu = 1$ and $\mu/ \alpha = 1$, $k\in[0.005,0.1]$ (yellow to blue). \textbf{Right}: Imaginary parts of the frequency as functions of $k/T$.}
\label{FIG.1}
\vspace{0.3cm}
\end{figure*}

As was pointed out in the previous section, when we set $\lambda=0$ and $\mathcal{J}\neq0$ in the bulk, the translations are spontaneously broken. For this case, the blackening factor and the black hole temperature simply reduce to
\begin{equation}
f(u)=1-\frac{u^{3}}{u_{h}^{3}}-\frac{\mu^{2} u^{3}}{4 u_{h}}\left(1-\frac{u}{u_{h}}\right),\qquad T=\frac{3}{4 \pi u_{h}}-\frac{\mu^{2} u _{h}}{16 \pi},
\end{equation}
which is exactly the same as the RN black hole. Based on this, QNMs of the fluctuating fields, $\delta a_{x}, \delta h_{t x}, \delta h_{x y}, \delta h_{u x}$ and $\delta \psi^{x}$ can be computed numerically and all the gapless modes have been plotted in Fig.\ref{FIG.1}.

From the numeric data, there are two gapless modes: 
The first one is a purely diffusive mode, $\omega=-iD_{\perp}k^2$, where one can easily check the diffusion constant is controlled by the shear viscosity in the way that $D_{\perp}=\frac{\eta}{\chi_{PP}} $ with the momentum susceptibility $\chi_{PP}=\epsilon+p=\frac{3}{2}\epsilon$. Here, $\epsilon$ and $p$ are the energy density and thermodynamic pressure. Obviously, this mode is associated with the conserved total momentum and the value of $D_{\perp}$ perfectly matches with that in the RN black hole since the gauge-axion coupling modifies neither $\eta$ nor $\chi_{PP}$. On top of this, there is an extra gapless mode that has 
a trivial dispersion relation, $\omega=0$. This mode is however absent in the RN case and should be viewed as the direct consequence of the symmetry breaking.

To manifest the existence of such a mode more, we compare our numeric result from holography with the prediction from viscoelastic hydrodynamics that is given by \cite{Amoretti:2019cef}
\begin{equation}\label{dispersion}
\omega_{\pm}=-\frac{i}{2}k^2\left(\xi_{\perp}+\frac{\eta}{\chi_{PP}}\right)\pm k\sqrt{\frac{G}{\chi_{PP}}-\frac{k^2}{4}\left(\frac{\eta}{\chi_{PP}}-\xi_{\perp}\right)^2}+\dots,
\end{equation}
where the shear Goldstone diffusion can be written as
\begin{equation}\label{gdiff}
\xi_{\perp}= G X, 
\end{equation}
with $X\equiv \lim_{\omega\rightarrow 0}\,\omega \,\text{Im} \mathcal{G}^R_{\mathcal{O}^x\mathcal{O}^x}(\omega,k=0)$ which in holography can be expressed as an analytic function of horizon and UV data \cite{Amoretti:2019cef}. In the regime where $G$ is sufficiently small comparing with $k$, the two gapless modes get decoupled as follows,
\begin{equation}
\omega_+=-i\frac{\eta}{\chi_{PP}}k^2+\mathcal{O}(iG),\,\,\,\,\,\omega_-=-i\xi_{\perp}k^2-\mathcal{O}(iG).
\end{equation}
In fluids, the shear modulus $G$ is exactly vanishing. Then, the decoupling limit above can always be reached for arbitrary value of $k$. As a result, the fluidity kills not only the real parts of the dispersion relations (\ref{dispersion}) but also  the imaginary part of the $\omega_{-}$ mode in the way that $\xi_{\perp}\rightarrow0$ as $G\rightarrow0$, which completely recovers the numeric result of the holographic model.\footnote{Note that the derivation of the hydrodynamic result (\ref{dispersion}) is in principle based on the assumption of low frequencies and small momenta and valid to order $k^2$.  However, our numeric result shows that it still works very well for rather large values of $k$.} Since it is a directly consequence of the spontaneous symmetry breaking, we call the gapless mode $\omega_-=0$ transverse Goldstone mode in contrast to \cite{Baggioli:2019abx}. 

Unlike acoustic phonons, such a transverse excitation with zero gradient energy obeys the equation of motion of a quantum mechanical free particle (instead of an oscillator) whose general solution is linear in time. Thus, this can be viewed as the linearized limit of a vortex in constant rotation \cite{Endlich:2010hf}. For this reason, one can refer this kind of modes as ``infinitesimal vortices" \footnote{Vortices in superfluids are gapped. On the contrary, the vortex configurations in ordinary fluids can be arbitrarily mild, involving only infinitely low momenta, which implies that these degrees of freedom should be included in the low energy description together with other gapless modes.}. Since they do not feature wave or diffusive behaviors, as was verified in  \cite{Gripaios:2014yha} with the field theory method that they do not appear as poles in correlators of physical observables. Moreover, the impact of the vortex on charge transport was also be investigated in holographic models with the same gauge-axion coupling \cite{Liu:2022bam}.

Another interesting issue is to check whether the diffusion constant $\xi_{\perp}$ obeys the lower bound that was proposed by Hartnoll earlier \cite{2015}. Obviously, it will strongly violate the bound when we choose the butterfly velocity as the characteristic velocity
\cite{Blake:2017qgd}, in contrast to the crystal diffusion in the longitudinal channel   \cite{Baggioli:2020ljz}. However, the bound does work if the speed of sound (which is zero here) is used to bound the diffusion.
\subsection{Pseudo-Goldstone mode}
Now, let us consider how the two gapless modes behave when we turn on an external source that breaks translations explicitly but weakly which amounts to set $\mathcal{J}\gg\lambda \neq0$ in the bulk. From the field theory perspective, once the spontaneously broken symmetry is approximate, the would-be massless excitation gets slightly gapped which is now called pseudo-Goldstone. In the presence of weakly explicit breaking, the low energy poles at zero momentum can be derived using the hydrodynamic formulation \cite{Kats:2007mq,Delacretaz:2017zxd},
\begin{equation}\label{18}
(\bar{\Omega}-i \omega)(\Gamma-i \omega)+\omega_{0}^{2}=0,
\end{equation}
where $\Gamma$ is the rate of the momentum relaxation and $\bar{\Omega}$ denotes the phase relaxation whose origin was widely discussed in recent studies of holography \cite{Donos:2021pkk,Ammon:2021slb} as well as effective theories \cite{Baggioli:2020nay,Baggioli:2020haa,Delacretaz:2021qqu,Armas:2021vku} and $\omega_{0}$ is called the pinning frequency related to the mass of the pseudo-Goldstone.\footnote{In viscoelastic hydrodynamics, $\omega_0^2=\frac{G}{\chi_{PP}}m^2$, where $m$ is the ``bare mass" of the pseudo-Goldstone in its Lagrangian \cite{Delacretaz:2017zxd}. However, this formula seems not working for fluids, as it has been clearly found in some previous work that the pinning effect also exist in these systems with a zero $G$ \cite{Li:2018vrz,Baggioli:2019abx}.} Solving Eq. (\ref{18}) we obtain a pair of modes,
\begin{equation}\label{20}
\omega_{\pm}=-\frac{i}{2}(\bar{\Omega}+\Gamma) \pm \frac{1}{2} \sqrt{4 \omega_{0}^{2}-(\Gamma-\bar{\Omega})^{2}},
\end{equation}
which are associated with the relaxed momentum and Goldstone. 
 
\begin{figure}[H]
\vspace{0.3cm}
\centering
\includegraphics[width=0.4\textwidth]{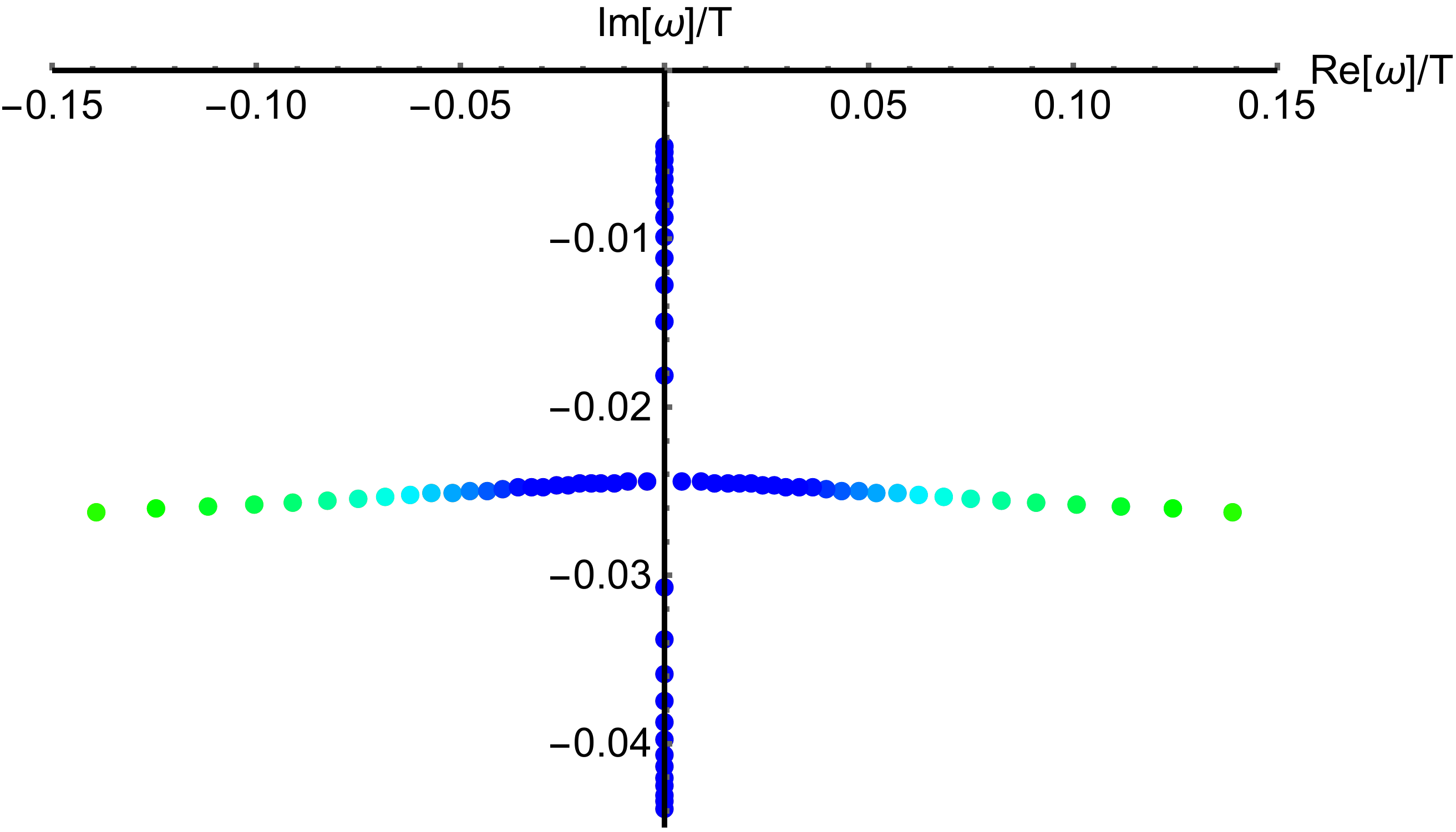}
\includegraphics[width=0.4\textwidth]{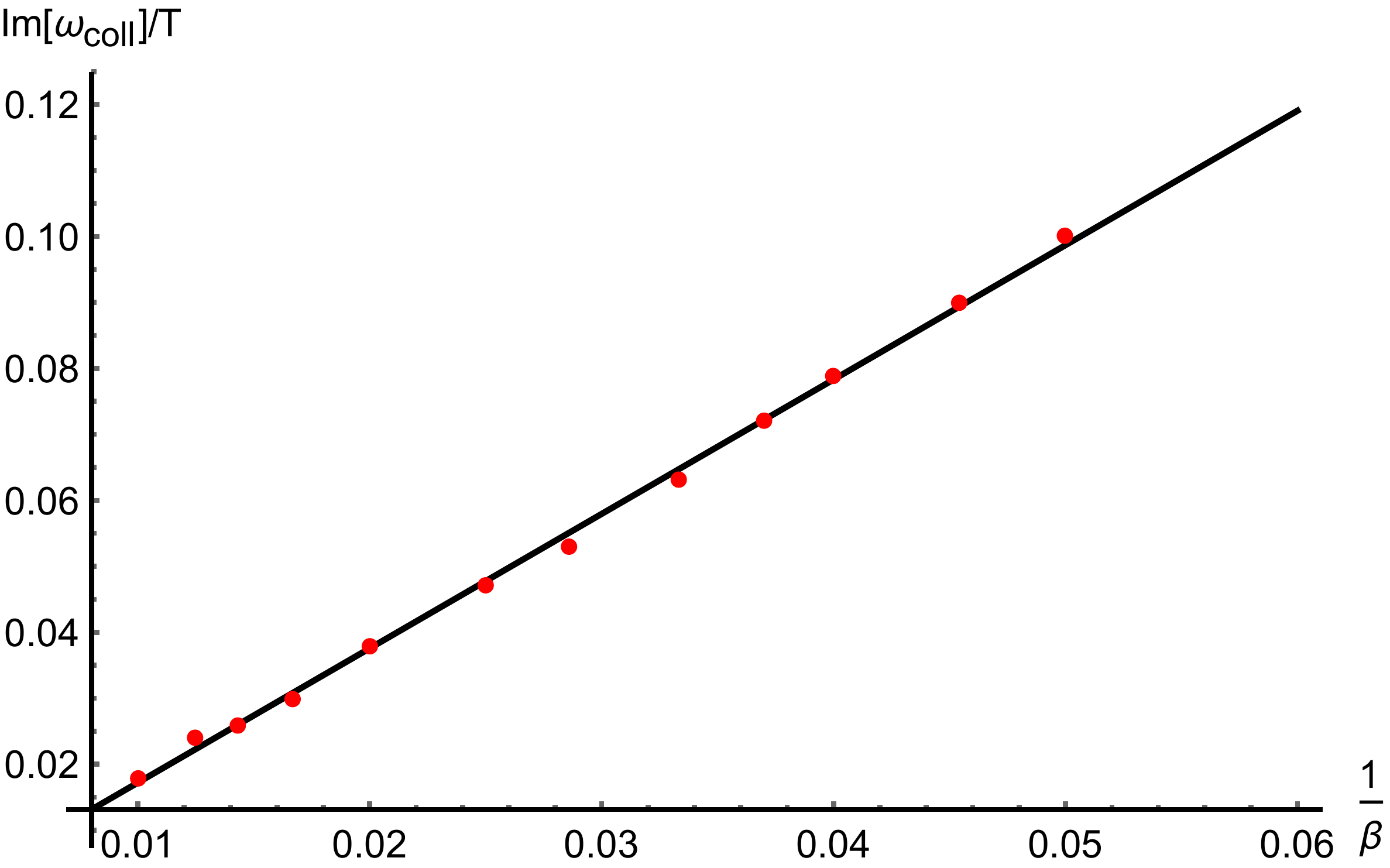}
\caption{\textbf{Left}: The motion of the lowest-lying modes on the complex plane. We fix $\mathcal{J} = 0.25$, $\lambda=0.003125$, $k=0$, $\alpha/ \mu = 1$ and dial $T / \mu \in[0.003,0.5]$ (green-blue). \textbf{Right}: The collision frequency as a function of $1 / {\beta}\equiv\left(\lambda/\mathcal{J}\right)^{\frac{1}{2}}$ where the red dots are the numeric data extracted from QNMs and the solid line is the fitting result. Here, we fix $T /\mu = 0.1$. When $\beta\gg 1$, the collision always happens in the hydrodynamic regime, i.e., $|\omega_{\text{coll}}|\ll T$. 
}\label{FIG.2}
\vspace{0.3cm}
\end{figure}

For $\omega_{0}^{2}<(\Gamma-\bar{\Omega})^{2}/4$, both of the two modes are damping, lying along the imaginary axis. At $\omega_{0}^{2}=(\Gamma-\bar{\Omega})^{2}/4 =0$, they collide on the imaginary axis, and then move off the axis when $\omega_{0}^{2}>(\Gamma-\bar{\Omega})^{2}/4$ acquiring a non-trivial real part of the frequency in their dispersion relations. In the last case, we find massive excitations in the spectrum. The behavior described in (\ref{20}) is in perfect agreement with what we see in the holographic model. The details have been exhibited in Fig.\ref{FIG.2}. 

\begin{figure*}[htbp]
\vspace{0.3cm}
\centering
\includegraphics[width=0.4\textwidth]{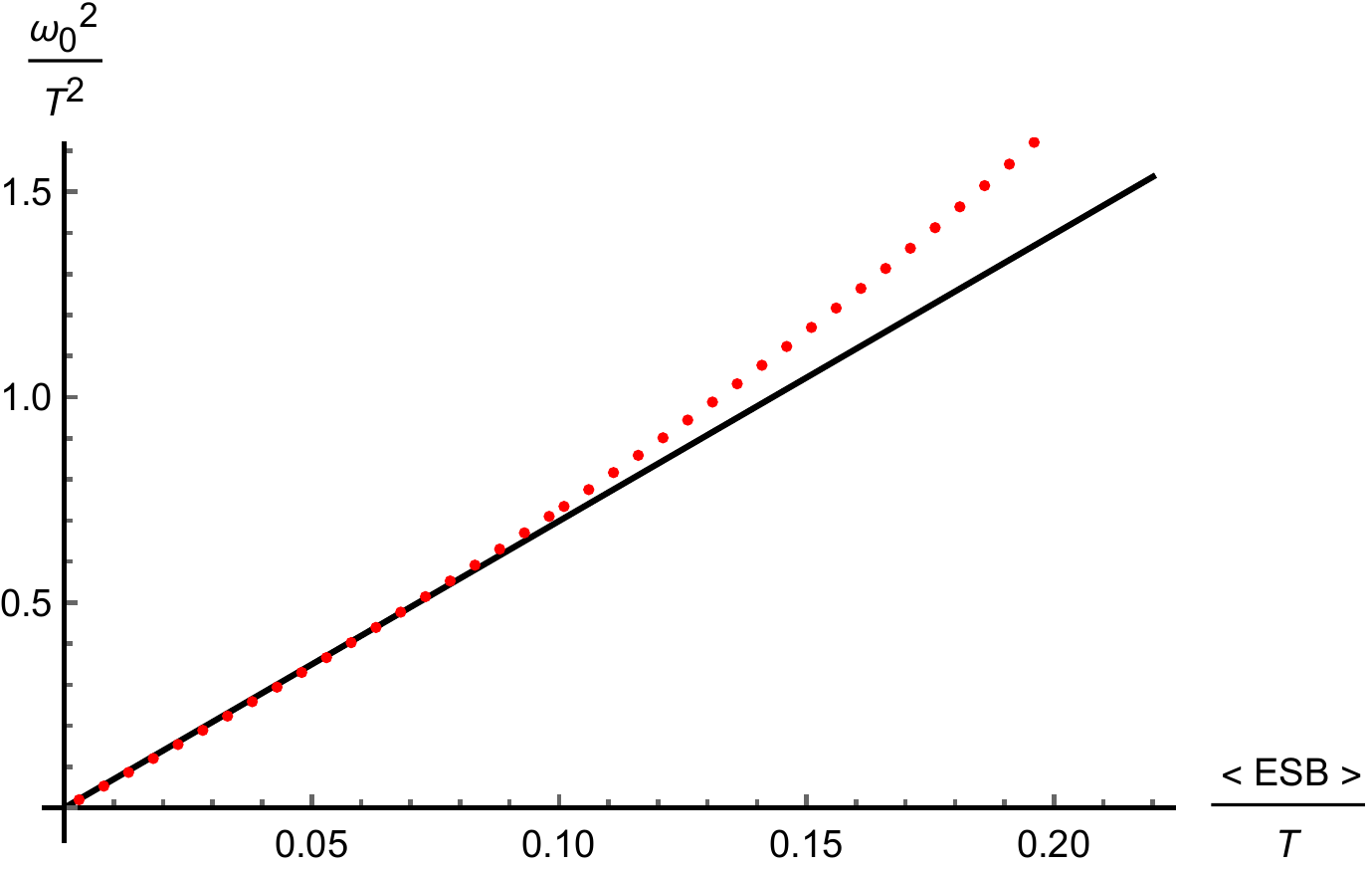}
\includegraphics[width=0.4\textwidth]{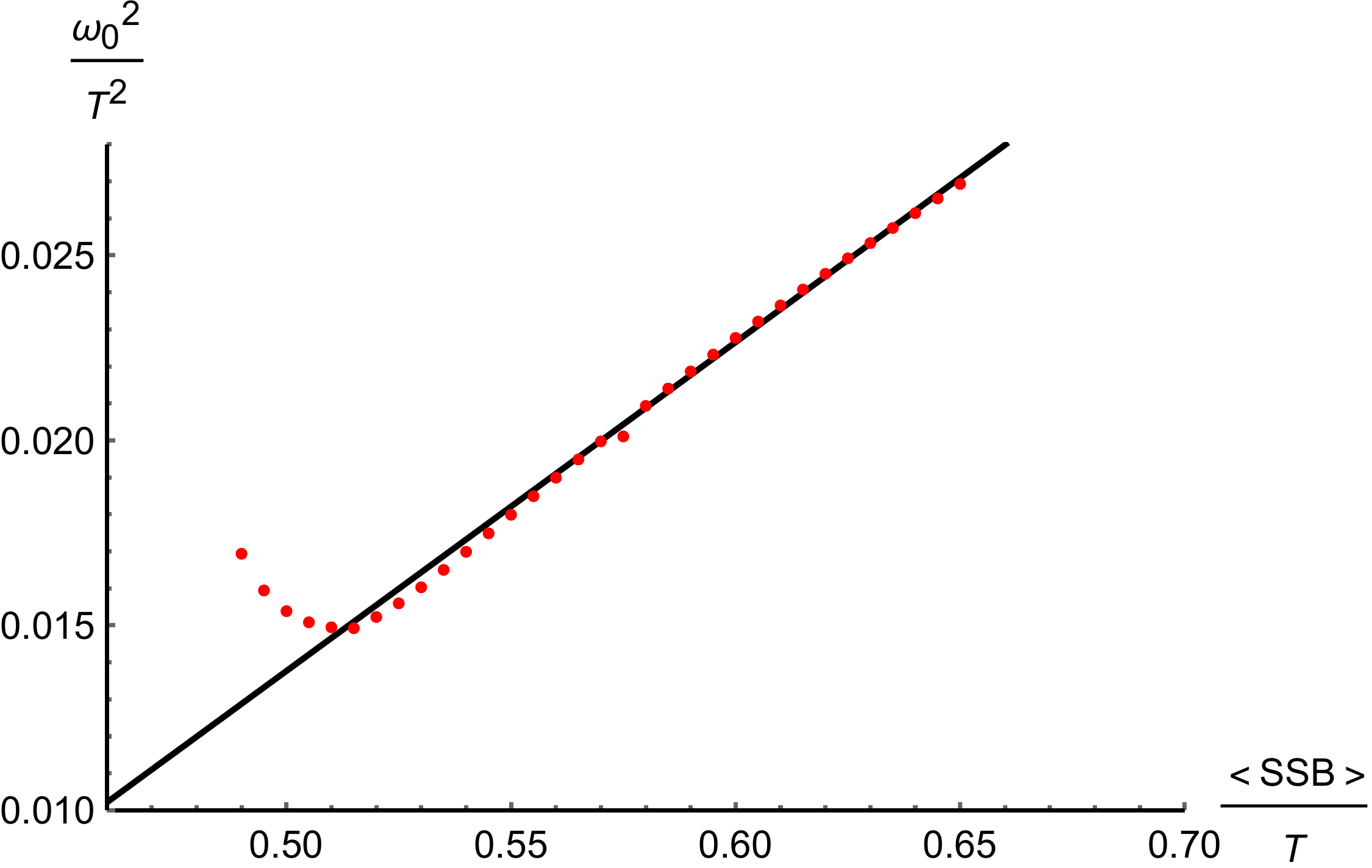}
\caption{Dependence of $\omega_0^{2} /T^2$ on $\langle\text{ESB}\rangle/T$ and $\langle\text{SSB}\rangle/T$. Left: We have set $\mathcal J$ = $0.25$ and $\lambda \in [0.00001, 0.01]$. Right: We have set $\lambda$= $0.0001$ and $\mathcal J \in [0.03,0.65]$. In both panels, we have fixed $\alpha/ \mu=1 $ and $T / \mu = 0.1$. }
\label{FIG.3}
\vspace{0.3cm}
\end{figure*}

\begin{figure*}[htbp]
\vspace{0.3cm}
\centering
\includegraphics[width=0.45\textwidth]{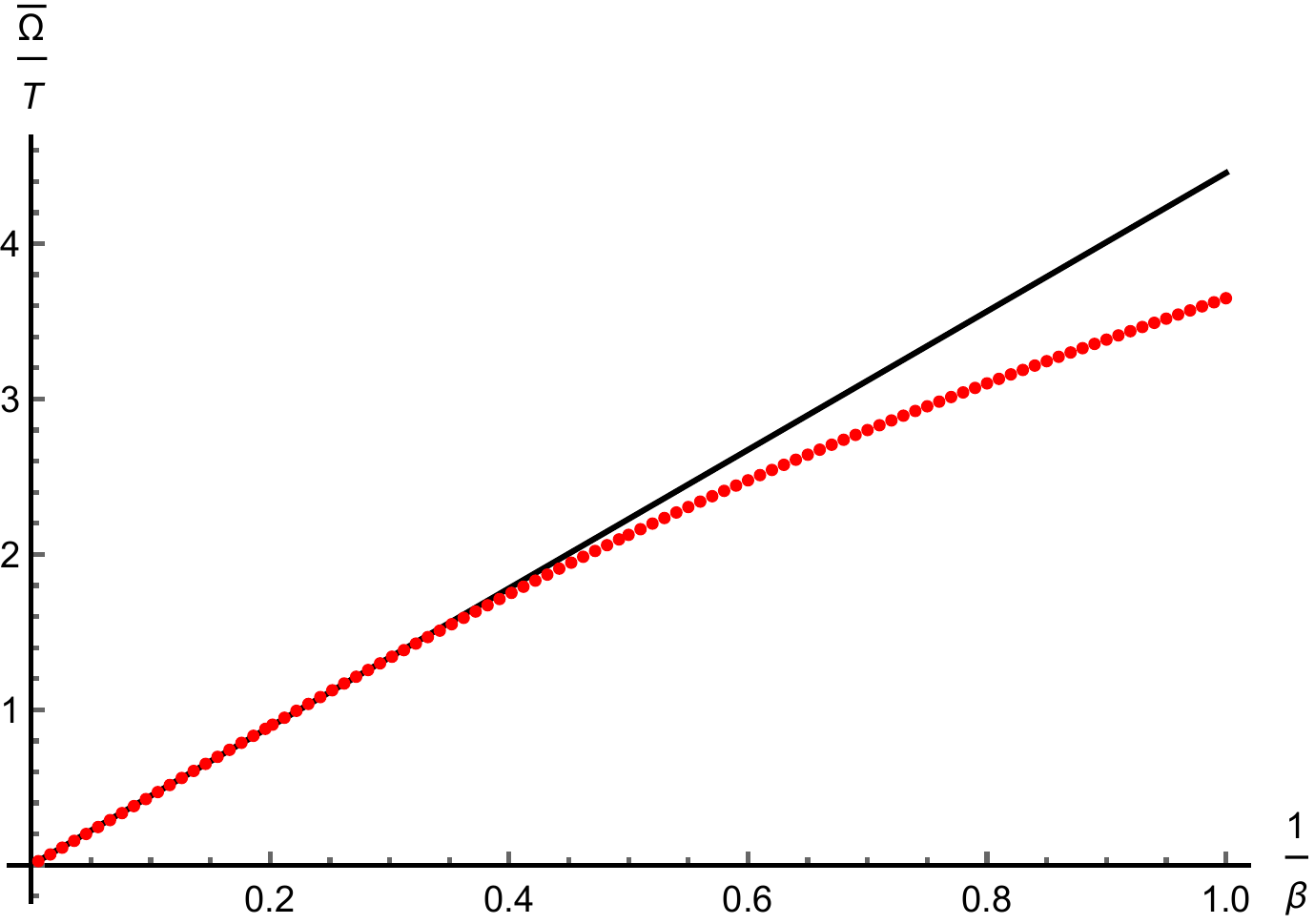}
\caption{Dependence of $\bar{\Omega} / T$ on $\beta$ for $T /\mu = 0.1$ and $\alpha/ \mu = 1$.   }
\label{FIG.4}
\vspace{0.3cm}
\end{figure*}

Since the three patterns above are determined from the competition of the three parameters  in (\ref{18}) and (\ref{20}), it is necessary to analyze how they depend upon the strength of SSB and explicit symmetry breaking(ESB). To do so, one can introduce two scales  $\langle\text{SSB}\rangle$ and $\langle\text{ESB}\rangle$ to characterize the two kinds of symmetry breaking just as in the solid case \cite{Ammon:2019wci}. The explicit breaking of translation should be related to the mass of the spin-1 graviton
\begin{equation}\label{20a}
m_{UV}^2=\lambda \,\alpha^2
\end{equation}
in the UV so that momentum is relaxed on boundary. Therefore, it is naturally to identify that
\begin{equation}
\langle \text{ESB}\rangle\equiv m_{UV}=\sqrt[] {\lambda} \alpha.
\end{equation}
In the weakly broken case, i.e., $\langle \text{ESB}\rangle \ll T$, the scaling of the momentum relaxation rate can be identified as \cite{Davison:2014lua,Hartnoll:2012rj,Lucas:2015vna}
\begin{equation}
\Gamma =\frac{\lambda \alpha^2}{4\pi T}\sim \frac{\langle\text{ESB}\rangle^2}{T}.
\end{equation}
Since the SSB is holographically realized by the $\mathcal{J}$ coupling, the associated scale can be defined as
\begin{equation}
{\langle \text{SSB}\rangle}\equiv\sqrt{\mathcal{J}} \alpha.
\end{equation}
In the presence of both ESB and SSB, one can introduce the following dimentionless parameter for simplicity,
\begin{equation}
\beta\equiv\frac{\langle \text{SSB}\rangle}{\langle \text{ESB}\rangle}=\left(\frac{\mathcal{J}}{\lambda}\right)^{\frac{1}{2}},
\end{equation}
representing the ratio of the two scales. Then, the pseudo-spontaneous breaking regime implies that $\beta \gg 1$. In this regime, we find the following scaling of the collision point(see the right panel of Fig.\ref{FIG.2}),
\begin{equation}
\frac{|\omega_{\text {coll}}|}{T} \propto \frac{1}{\beta}\ll 1.
\end{equation}
Since we set a small $\lambda$ so that $\Gamma$ is sufficiently small comparing with the other parameters, we will neglect its effects and extract $\omega_{0}$ and $\bar{\Omega}$ from (\ref{20}) by fitting with the numeric data. The dependence of $\omega_{0}$ on $\langle \text{ESB}\rangle$ and $\langle \text{SSB}\rangle$ has been plotted in Fig.\ref{FIG.3} and shows that
\begin{equation}
\omega_{0}^{2} \propto\langle \text{ESB}\rangle\langle \text{SSB}\rangle,
\end{equation}
reminiscent of the Gell-Mann–Oaks–Renner(GMOR) relation relation firstly discovered in QCD \cite{GellMann:1968rz}, and also has been touched in some studies of holography recently \cite{Andrade:2017cnc,Wang:2021jfu}.
Moreover, the scaling of the phase relaxation 
\begin{equation}
\frac{\bar{\Omega}}{T} \propto \frac{\langle \text{ESB}\rangle}{\langle \text{SSB}\rangle}=\frac{1}{\beta},
\end{equation}
has also been shown in Fig.\ref{FIG.4}. Then, in the pseudo-spontaneous regime with $\beta\gg 1$, we have 
\begin{equation}
\frac{\omega_{0}}{\bar{\Omega}} \propto \frac{\langle \text{SSB}\rangle}{T}\beta^{1/2}\gg 1
\end{equation}
which means the pinning frequency dominates the real part of (\ref{20}), therefore can be treated as the physical mass of the pseudo-Goldstone modes.

\subsection{Dispersion relation and k-gap}
\begin{figure}[H]
\vspace{0.3cm}
\centering
\includegraphics[width=0.4\textwidth]{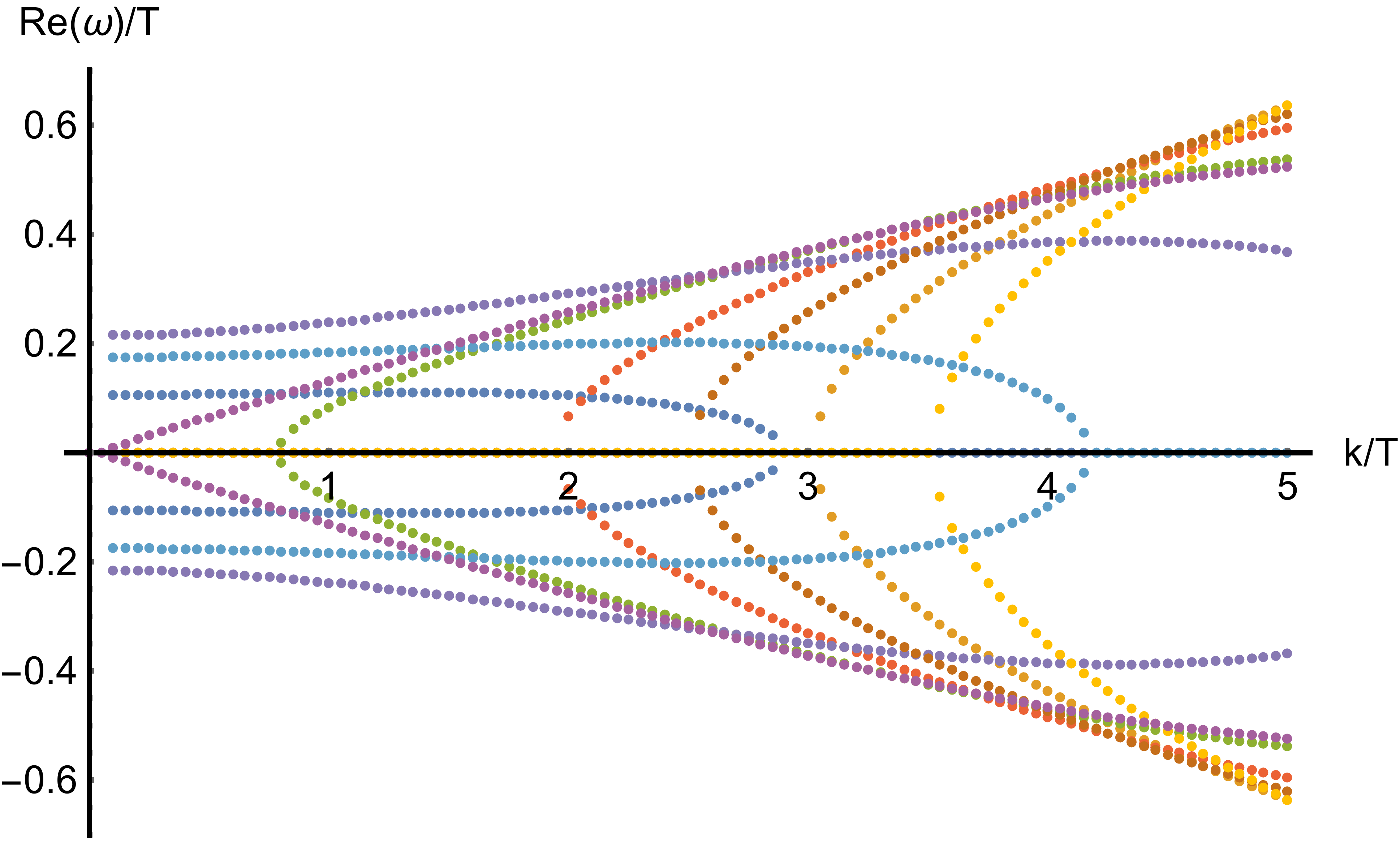}
\includegraphics[width=0.4\textwidth]{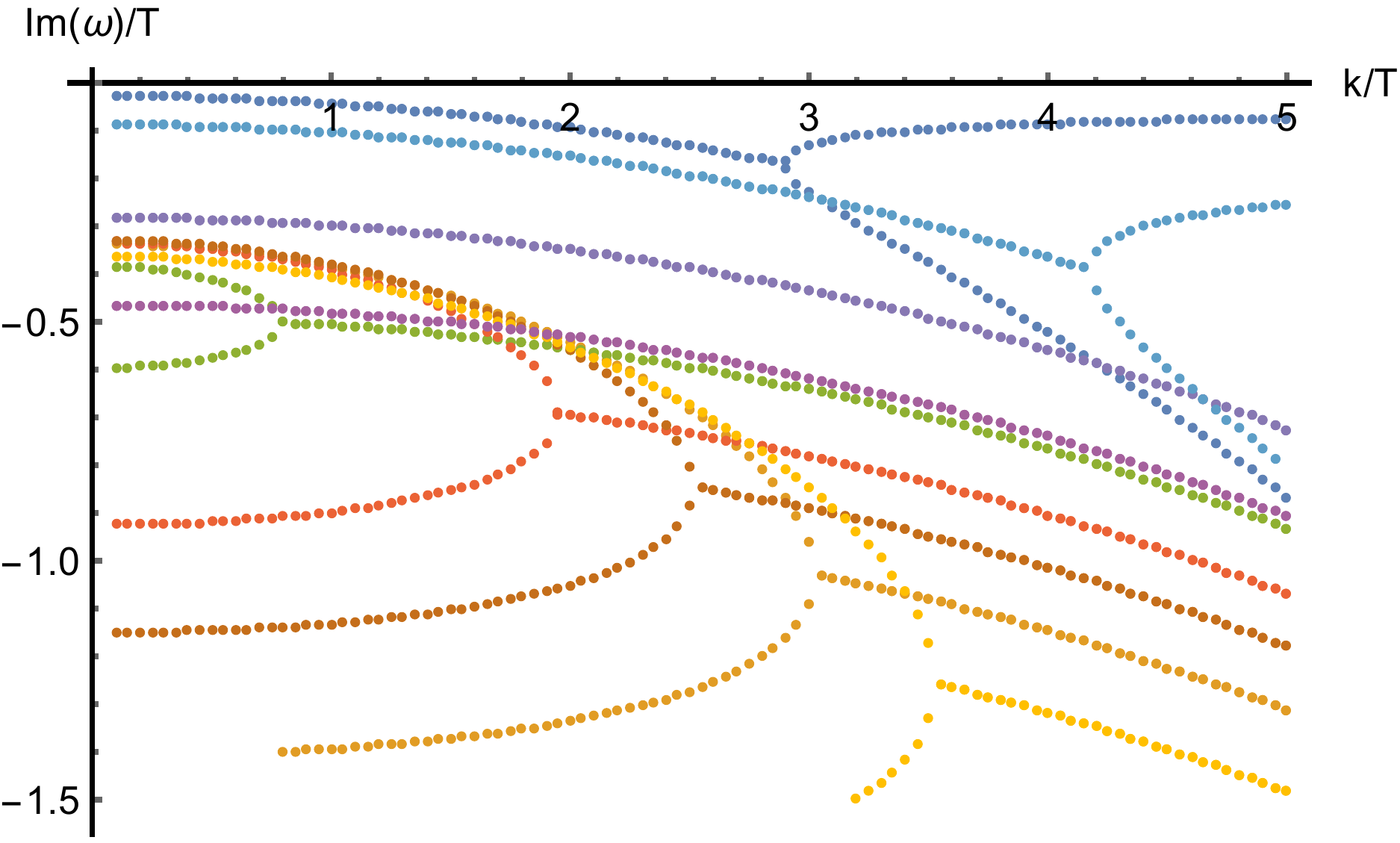}
\caption{Diperstion relations of the lowest-lying modes for $\mathcal{J} = 0.25$ , $\alpha = 1$,  $T / \mu = 0.1$ and $\lambda \in [0.0001, 0.15] $, i.e. $ \beta \in[ 0.65,50 ]$ (yellow-blue). The critical value of $\beta$ can be fixed as $\beta_c \approx 3.049$. 
}\label{FIG.5}
\vspace{0.3cm}
\end{figure}

In the final subsection, we will discuss the dispersion relations of the shear modes for finite momentum. The numeric result has been displayed in Fig.\ref{FIG.5}. We also start the dicussions with the pseudo-spontaneous case, i.e., supposing $\beta$ is sufficiently large.  We find that the dispersion relations of pseudo-Goldstone modes for small $k$ are given by
\begin{equation}
\omega_{\pm}\approx \pm  \sqrt{ \omega_{0}^{2}-\frac{(\Gamma-\bar{\Omega})^{2}}{4}}-i\left(\frac{\bar{\Omega}+\Gamma}{2}+D_\perp k^2\right)+\dots
\end{equation}
in the pseudo-spontaneous regime. Apart from the real part, imaginary part has a pseudo-diffusive like scaling which implies that the pseudo-Goldstone is coupled with the momentum and may have a contribution to the transport. This is different from the purely SSB case where the Goldstone is completed frozen. When $k/T$ is large and increased,
these two modes move towards the imaginary axis of the frequency, collide, split and move in the opposite directions on the imaginary axis again.

\begin{figure}[H]
\vspace{0.3cm}
\centering
\includegraphics[width=0.45\textwidth]{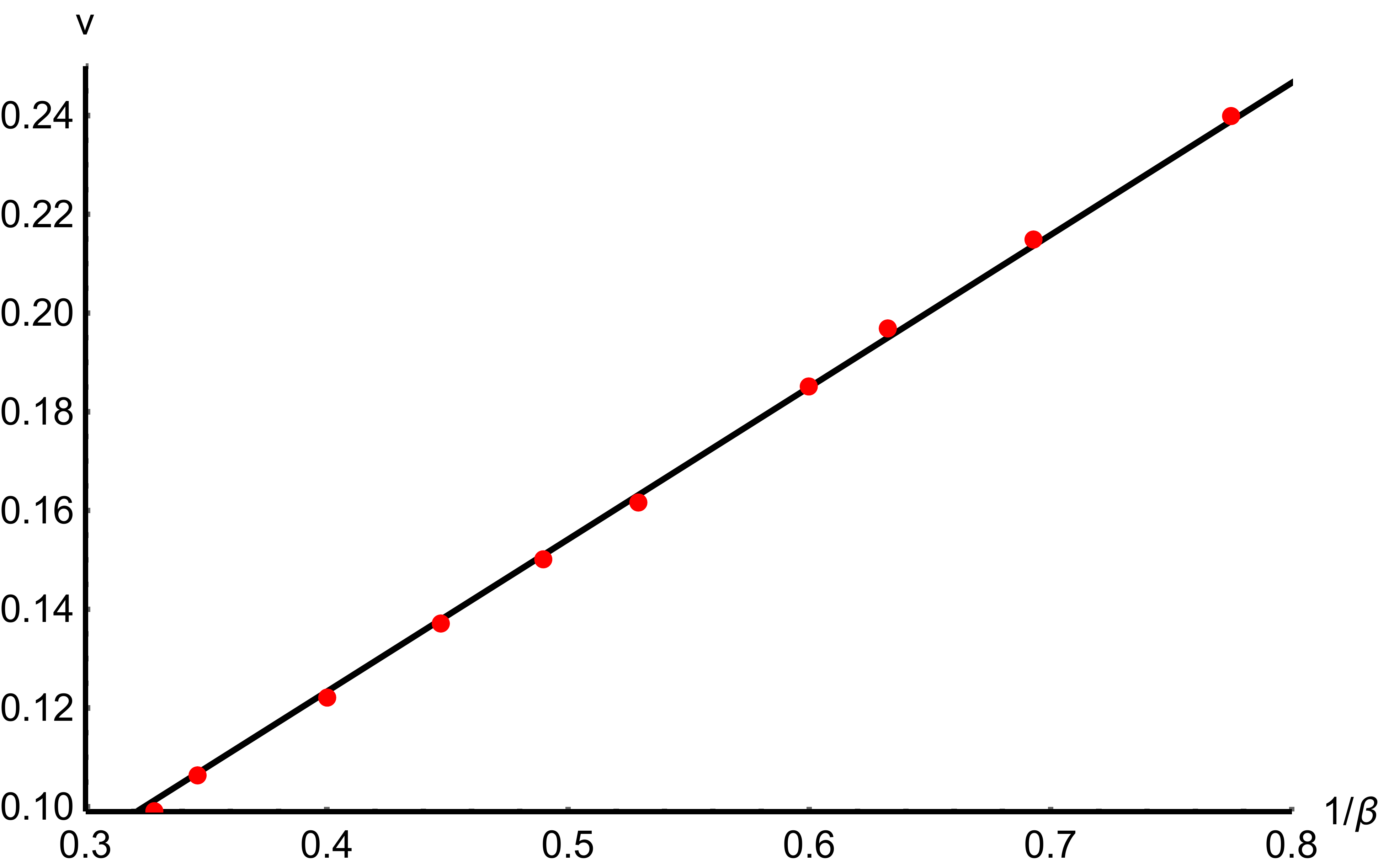}
\includegraphics[width=0.45\textwidth]{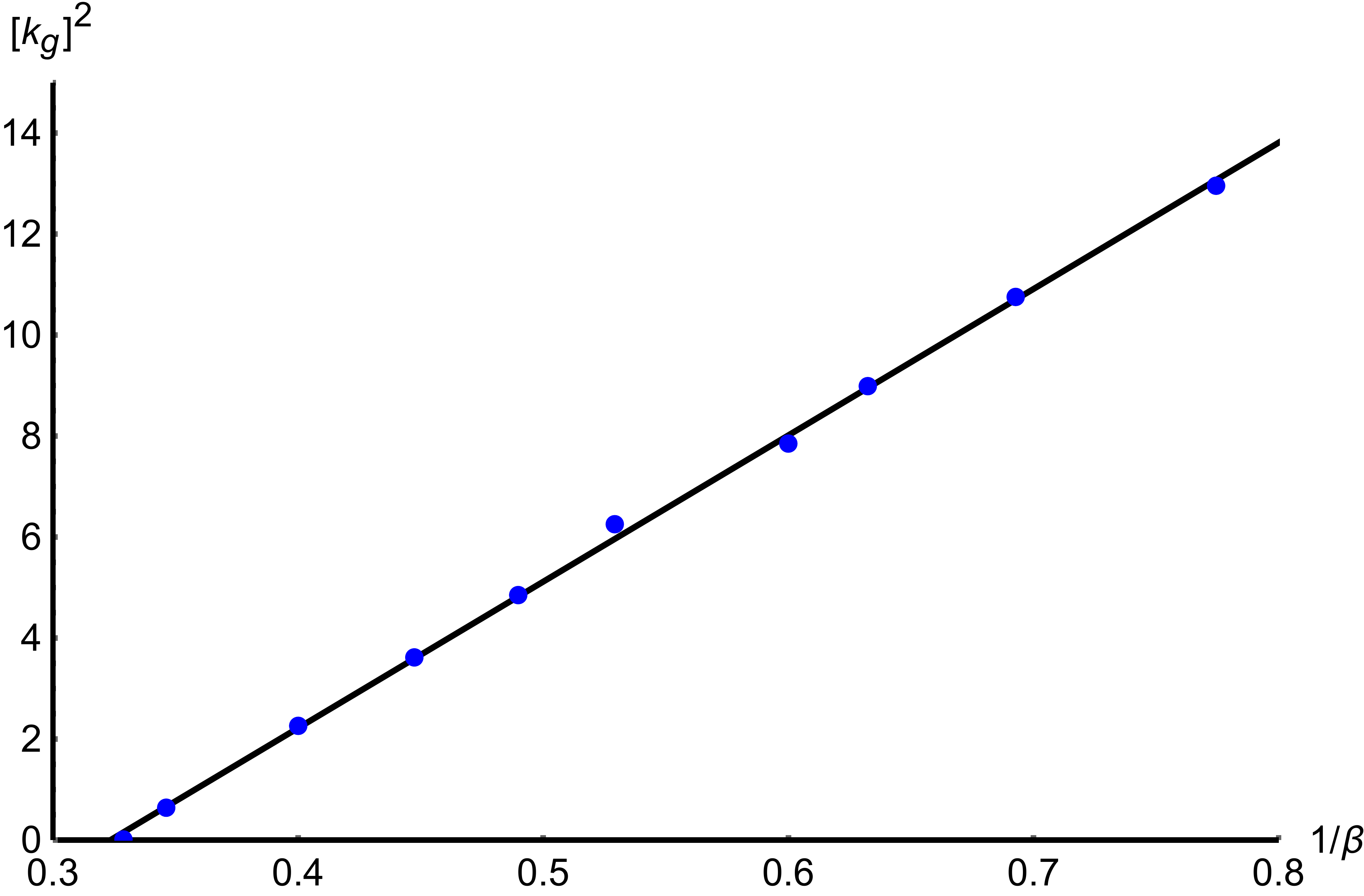}
\caption{The scalings of $v$ and $k_g$. Here, we have set $T /\mu = 0.1$, $\alpha = 1$ and $\mathcal{J} = 0.25$ and $ \beta \in[0.65,50]$.}
\label{FIG.6}
\vspace{0.3cm}
\end{figure}

Moreover, if we increase $\langle\text{ESB}\rangle$ more and more, the pseudo-spontaneous breaking pattern breaks down duo the significant effect of $\Gamma$. Meanwhile, the mass of the pseudo-Goldstones get smaller and smaller. When the value of $\beta$ is down to certain critical value $\beta_c$, two sound modes emerge. These two modes are, nevertheless, a result of fine-tuning. Once we increase $\langle\text{ESB}\rangle$ further, they will be destroyed and become pseudo-diffusive at low momenta and propagating at high momenta which is similar as the purely explicit breaking pattern \cite{Davison:2014lua,Kim:2014bza}. In this case, the associated value of momentum $k_g$ that separates the two behaviors is called a k-gap.
Scalings of the emergent sound speed $v$ and $k_g$ have been fixed by the data in Fig.\ref{FIG.6},
\begin{equation}
v\sim \frac{\langle\text{ESB}\rangle}{\langle\text{SSB}\rangle},\quad k_g^2\sim \frac{\langle\text{ESB}\rangle}{\langle\text{SSB}\rangle}.
\end{equation}
However, the origin of such an emergent sound speed due to the ESB of translations still remains mysterious.

\section{Conclusions}\label{section4}
In this paper, we investigate low energy transverse excitations of a simple holographic fluid model with spontaneous broken translations. In the purely spontaneous breaking case, there are two decoupled gapless shear modes one of which is diffusive and associated with the conserved momentum. The other is the Goldstone mode of the symmetry breaking that behaves like gapless vortices. 

Then, we switch on an external source to break the translations also explicitly. When the effect of momentum relaxation is weak, the would-be gapless modes are coupled and form a pair of slightly gapped pseudo-Goldstone modes via colliding on the imaginary axis of frequency. We verify that the mass of the pseudo-Goldstone modes satisfies the GMOR relation. Their lifetime is characterized by the phase relaxation which comes from breaking simultaneously the translations and global internal shift symmetry \cite{Ammon:2021slb}. When the momentum relaxation is strong enough, we find that the picture of pseudo-Goldstones breaks down and the mass of the shear modes becomes smaller as the increase of the relaxation rate. Finally, they become massless and generate 
a $k$-gap that sets a (pseudo-)diffusion-to-sound crossover which is similar as the purely ESB case. 
\subsection*{Acknowledgments} 
We would like to thank Matteo Baggioli for numerous helpful discussions on (pseudo-) Goldstone modes in fluids and Xi-Jing Wang for sharing his experience in numerical calculations. This work is supported by NSFC No.11905024.

\bibliographystyle{apsrev4-1}
\bibliography{references}
\end{document}